# Fusion research using Azure A100 HPC instances

Igor Sfiligoi
University of California San Diego
La Jolla, CA, USA
isfiligoi@sdsc.edu

Jeff Candy
General Atomics
La Jolla, CA, USA
candy@fusion.gat.com

Devarajan Subramanian
Drizti Inc
Toronto, ON, Canada
devs@drizti.com

*Abstract*— Fusion simulations have in the past required the use of leadership scale HPC resources to produce advances in physics. One such package is CGYRO, a premier multi-scale plasma turbulence simulation code. CGYRO is a typical HPC application that would not fit into a single node, as it requires O(100 GB) of memory and O(100 TFLOPS) worth of compute for relevant simulations. When distributed across multiple nodes, CGYRO requires high-throughput and low-latency networking to effectively use the compute resources. While in the past such compute may have required hundreds, or even thousands of nodes, recent advances in hardware capabilities allow for just a couple of nodes to deliver the necessary compute power. This paper presents our experience running CGYRO on NVIDIA A100 GPUs on InfiniBand-connected HPC resources in the Microsoft Azure Cloud. A comparison to older generation CPU and GPU Azure resources as well as on-prem resources is also provided.

## I. Introduction

Fusion energy research has made significant progress over the years, yet the complexity of the turbulence in toroidal plasmas makes it difficult to accurately predict fusion reactor performance. While experimental methods are essential for gathering new operational modes, simulations are used to validate basic theory, plan experiments, interpret results on present devices, and ultimately to design future devices. The tested Fusion simulation tool was CGYRO [1], an Eulerian gyrokinetic solver designed and optimized for collisional, electromagnetic, multiscale simulation, which is widely used in the Fusion research community.

CGYRO simulations have traditionally been executed on leadership-class HPC centers, such as NERSC Cori, OLCF Summit and TACC Stampede2. While leadership-class HPC centers provide for a very pleasant and effective work environment, they are heavily sought after and typically over-subscribed. Cloud computing has been recently identified as a viable alternative [2] and in this paper we update that assessment by providing benchmark information for the recently added NVIDIA A100-based Microsoft Azure Cloud HPC instances.

One of the limiting factors of Cloud HPC resources is the lack of native HPC work environments. In order to make CGYRO accessible to a typical Fusion scientist, a turn-key solution is essential. The HPCBOX [3] team provided us with such a solution, and all the tests were run in that environment.

## II. Azure ND A100 V4 Instances

In 2021 Microsoft released the NVIDIA A100 GPU-providing resources as ND A100 v4 instances. Each comes with 8 NVLINK-connected GPUs, 96 CPU cores and 8 200Gbps InfiniBand (IB) ports. At least on paper, these resources are an ideal platform for CGYRO simulations. Unfortunately, the availability of these resources is still very constrained, and it was difficult to obtain more than 16 instances at any point in time. It should be noted that getting resources in spot mode is typically easier than on-demand, due to the difficulty of raising the quota limits, so we extensively used those in our tests. HPCBOX supports spot instances out of the box, and it also automatically restarts preempted simulation jobs from the last checkpoint, if needed.

According to Azure documentation [4], each GPU is paired with its own InfiniBand port. Unfortunately, this topology is not exposed inside the instance, so the MPI layer has no way to auto-detect the proper routing. Microsoft does however provide a recipe for manually tuning NCCL-based applications, so one can use that documentation to infer the expected configuration, which turns to be simply matching each GPU to the IB port in the same order.

Our previous experience [2] showed that the NVIDIA-provided OpenMPI, bundled with the NVIDIA HPC SDK, generally delivers the best performance for CGYRO. Unfortunately, that version does not allow for manual pairing of GPUs to IB ports, which resulted in sub-optimal network pairing, i.e., treating all IB ports as equivalent. This results in a large fraction of the networking traffic to also traverse the PCIe and CPU-to-CPU links, making those the communication bottleneck.

OpenMPI 4.1.1 does allow manual tunning, so we also manually compiled it with the NVIDIA HPC SDK and tied each GPU to its own IB port there. Unfortunately, this version seems to have an upper limit on the number of processes per GPU that it can handle, i.e., 4 processes per GPU and 32 processes per node. This prevented us from optimally placing the CGYRO's MPI processes on the nodes. Nevertheless, the improved networking routing provided a significant boost in performance in certain configurations, as described in the next section.

In the end we put in place two MPI configurations for users to choose from. We plan to engage with both MPI providers and Microsoft in the future to improve on the situation.

This work was supported by the U.S. Department of Energy under awards DE-FG02-95ER54309 (General Atomics Theory grant), DE-SC0017992 (AToM SciDAC-4 project).

Pre-print version of the accepted contribution to SC21

## III. BENCHMARK RESULTS

CGYRO can be used to run simulations at multiple scales, but currently the most useful, medium-scale ones are well represented by the "nl03" benchmark case. At the top level, the nl03 benchmark case rigidly splits the base problem into 64 sub-problems, based on the detector's toroidal angle, and each of those sub-problems can be optionally further split across the other dimensions. Due to the nature of the compute algorithm [5], the most MPI-intensive communication is between processes working on the same slice of the top sub-problem, so packing 64 MPI ranks on a single node is highly desirable. While multiple MPI ranks can share a single GPU, within reason, a single CPU core should not be shared, for performance reasons. This was a problem with previous generation Azure GPU instances, the V100-based ND v2, which had only 40 CPU cores, but it is doable on the newer A100-based instances, which have 96 cores. Unfortunately, the better performing OpenMPI 4.1.1 does not support this setup, so we ran an additional test with just 32 MPI ranks per node.

The CGYRO nl03 benchmark case requires approximately 560 GB of GPU memory, which results in the need for at least 16 A100 GPUs, using about 35 GB out of the 40 GB that are available, i.e., at least two Azure instances. To evaluate scaling, we ran the simulation from 2 to 16 nodes using both the NVIDIA-provided OpenMPI and the tuned OpenMPI 4.1.1. The runtimes for a single benchmark step are presented in Table 1, alongside the time spent in communication alone.

TABLE I. CGYRO BENCHMARK RESULTS ON ND A100 V4

| Number of nodes | MPI version | MPI ranks per node | Total time | Communication time |
|---|---|---|---|---|
| 2 | NVIDIA | 64 | 200s | 166s |
| 2 | 4.1.1 | 32 | 265s | 224s |
| 4 | NVIDIA | 64 | 150s | 132s |
| 4 | 4.1.1 | 32 | 160s | 137s |
| 8 | NVIDIA | 64 | 117s | 103s |
| 8 | 4.1.1 | 32 | 100s | 88s |
| 16 | 4.1.1 | 32 | 67s | 60s |

As can be seen, the ability to run all 64 MPI ranks on a single node provides a significant advantage at low node counts, but the superior overall network performance of the tuned OpenMPI 4.1.1 more than compensates at higher node counts. We can only speculate how much faster would the compute be if we could have used 64 MPI ranks per node with OpenMPI 4.1.1.

To put the observed benchmark runtimes in context, Table 2 contains the results presented in our previous work [2]. As can be seen, 16 Azure ND A100 V4 instances outperform 16 OLCF Summit NVIDIA V100-based nodes, while 8 Azure ND A100 V4 instances outperform 64 NERSC Cori Intel KNL-based nodes. The ND A100 V4 instances are also about twice as fast as the previous-generation V100-based NDv2 instances at same instance count, and 16 ND A100 V4 instances outperform 36 AMD CPU-based HB v2 instances.

TABLE II. PAST CGYRO BENCHMARK RESULTS

| Provider | Resource type | Number of nodes | Total time | Communication time |
|---|---|---|---|---|
| OLCF | Summit V100 | 16 | 82s | 46s |
| NERSC | Cori KNL | 16 | 372s | 120s |
| NERSC | Cori KNL | 64 | 112s | 46s |
| Microsoft Azure | ND v2 (V100) | 4 | 397s | 293s |
| Microsoft Azure | ND v2 (V100) | 16 | 121s | 92s |
| Microsoft Azure | HB v2 (CPU) | 9 | 289s | 64s |
| Microsoft Azure | HB v2 (CPU) | 36 | 87s | 45s |

## IV. CONCLUSIONS

The recently added Microsoft Azure ND A100 V4 instances provide an appealing option for running medium-scale CGYRO fusion simulations in the Cloud. At small instance counts, these instances are faster than both OLCF Summit at the same node count and NERSC Cori at significantly higher node count.

Given that the availability of Azure ND A100 V4 instances is currently still very constrained, the leadership-class systems are still needed for any large-scale simulation. But for fusion scientists that have no access to on-prem HPC resources, the Cloud resources now provide a viable path forward for more ordinary simulation needs.

In an effort to simplify the transition from on-prem to Cloud resources, we performed all our tests using the HPCBOX solution. HPCBOX provided us with a fully configured system, including the provisioning layer, worker node Cloud images and the needed MPI libraries. It also includes a virtual desktop environment that should make it accessible to most science users.

ACKNOWLEDGMENT

This work was supported by the U.S. Department of Energy under awards DE-FG02-95ER54309 (General Atomics Theory grant), DE-SC0017992 (AToM SciDAC-4 project).